\documentclass[aps,letterpaper,prl,twocolumn,amsmath,amssymb]{revtex4}
\usepackage{exscale,latexsym}
\usepackage[dvips]{graphicx}
\begin{document}
\preprint{KLT}
\title{String theory and the mapping of gravity into gauge theory}
\author{N.~E.~J.~Bjerrum-Bohr}
\email{bjbohr@nbi.dk}
\affiliation{The Niels Bohr Institute,\\
Blegdamsvej 17, DK-2100 Copenhagen,
Denmark\footnote{\footnotesize{Present address.}}
\\and\\ Department of Physics,
University of California at Los Angeles,
Los Angeles, CA 90095-1547}
\date{\today}
\begin{abstract}
The relationship between on-shell tree level scattering amplitudes of
open and closed strings, discovered some time ago by Kawai, Lewellen
and Tye, is used at field theory level (at $O(\alpha'^3)$) to establish
a link between the general relativity and the non-abelian Yang-Mills
effective actions.
Insisting at the effective Lagrangian level that any tree $N$
point gravity on-shell scattering amplitude is directly factorisable
into a sum of $N$ point left-right products of non-abelian Yang-Mills
tree on-shell scattering amplitudes, non-trivial mappings of
the effective general relativity operators into the effective
non-abelian Yang-Mills operators are derived.
Implications of such mapping relations of the field operators are discussed.
\end{abstract}\maketitle

\section{Introduction}
One of the interesting aspects of modern string theory is that it enables
the conjecturing of links of a deep and profound nature, between seemingly
unrelated field theories.

Here the string relations, which factorizes any $N$ point scattering
amplitude of a closed string mode into a product of $N$ point scattering
amplitudes of left-right propagating modes of open strings, will be used
to link gravity and gauge theory.
The open-closed string relations discussed here are the Kawai-Lewellen-Tye
relations~\cite{KLT}.
As will be observed they impose a very non-trivial relationship between the
coefficients of on one side the generic general relativity effective
Lagrangian, and on the other side the general effective non-abelian
Yang-Mills massless vector field Lagrangian. The lowest order terms
were discussed previously in~\cite{Bern3}.

Gravity and gauge theories are two fundamental theories in the
understanding of high energy physics, both based on the principles of
local gauge invariance, but despite that they have
rather dissimilar field dynamics. While general relativity is only
renormalizable in an effective manner~\cite{Weinberg}, including at
each loop order new invariant field operators into the theory, a non-abelian
Yang-Mills gauge theory is a completely renormalizable theory in four
dimensions.
Besides that, the theory of general relativity, being an
infrared safe theory, does not share the UV asymptotic freedom associated
with non-abelian Yang-Mills gauge theories. The KLT connections impose
therefore a mysterious relationship between rather incongruent theories.

The KLT-relations can directly be used as an efficient way of calculating
$N$ point gravity amplitudes from $N$ point gauge
amplitudes at tree level~\cite{Berends}. By $D$-dimensional unitary
cuts of loop diagrams, i.e. using the analytic properties of the S-matrix,
also gravity loop amplitudes can be related to gauge theory ones,
through the KLT-relations~\cite{Bern11,Bern1,Bern2}. See
also~\cite{Bern0,Dunbar}. For a recent detailed review of the
applications of the KLT-relations, see~\cite{Bern2b}.
The KLT-relations do not hold only for the pure fields of the two
theories, but also in the presence of matter, as shown in~\cite{Bern4}.

Evidently this emphasizes that the KLT-relations are a rather useful and
labor conserving tool, which can be used in calculations of scattering amplitudes
in gravity. This has been used, for example, to demonstrate that $N=8$
supergravity is less divergent in the ultraviolet than had been appreciated~\cite{Bern1}.
However the intriguing question naturally arises, if the KLT-relations have
a role, in a deeper interplay between general relativity and and
non-abelian Yang-Mills theory.

The KLT-relations provide a hope for a more general link between the two
theories, than just for on-shell tree amplitudes. This can compliment the
Maldecena conjecture~\cite{Maldecena} which provides a relation between
strong and weak phases of the theories.
A natural place to begin such an investigation of the KLT-relations is
obviously at the Lagrangian level. However comparison of the effective
Lagrangians of the two theories, yields no immediate factorization of
gravity into a ``square'' of an Yang-Mills theory~\footnote{\footnotesize
Some initial steps in the quest for a factorization at the Lagrangian
level has been made by~\cite{Siegel}}.
An obvious complication of such investigations is, that the S-matrix
scattering amplitudes which go into the KLT-relations are on-shell.
Furthermore, string theory does not contain a Lagrangian formalism.
In~\cite{Bern3} it was shown that the tree level properly
regularized gravity vertices is reconstructible from the Yang-Mills
amplitudes for certain choices of the non-abelian gauge.

In this letter the two field theories will be compared at the
Lagrangian level. Through the comparison of the scattering amplitudes
and use of the KLT-relations, it will be seen how the coefficients of the
two effective Lagrangians have to be related. This does not solve the
fundamental problem of understanding the interplay between gravity and
gauge theory, seemingly given by the KLT-relations, but it provide means for
further investigations, and presents some rather astonishing links
between the field operators of non-abelian Yang-Mills theory and gravity.

The structure of this letter will be as follows. First it will be showed
how to construct the effective Lagrangians
in general relativity and Yang-Mills theory. Here it is very labor
conserving to determine early which field operators are irrelevant in
the process of comparing the scattering amplitudes. Only operators which do
not vary under a general field redefinition will
be included in the generic Lagrangians.
Next the explicit scattering amplitudes for the two theories based on
their respective generic Lagrangians
will be presented. The coefficients on the gravity side will be compared
with the coefficients on the Yang-Mills side,
insisting that the field theory limit of the KLT-relations have to hold.

Throughout this paper, units $c=\hbar=1$ and the metric
convention $(+---)$ will be employed.

\section{Effective actions for gravity and Yang-Mills}
Einstein's theory of general relativity tells us that the pure Lagrangian
for the gravitational field, in absence of a cosmological
constant, has to take the form
\begin{equation}
{\cal L} = \sqrt{-g} \bigg[{\frac{2R}{\kappa^2}}\bigg]
\end{equation}
where $g_{\mu\nu}$ denotes the gravitational field,
$g \equiv \det(g_{\mu\nu})$ and $R$ is the scalar curvature tensor.
$\kappa^2 \equiv 32\pi G$, where $G$ is the gravitational
constant. The convention,
$R^\mu_{\ \nu\alpha\beta} = \partial_\alpha \Gamma_{\nu\beta}^\mu - \partial_\beta \Gamma_{\nu\alpha}^\mu + \ldots$ is followed.

The possibility of including higher derivative contributions into the
action is not excluded by the principles of general relativity.
The implications at the classical level by the addition
of such terms into the action are very limited at
low energies~\cite{Stelle}.

It is a well known fact that the pure theory of Einstein gravity is
formally renormalizable at $D=4$ and one loop~\cite{Veltman}, but this
is a coincidence. Trivially it happens because the Ricci-tensor vanishes
on-shell for the pure theory. Already at 2-loop order the theory was shown
to be truly a non-renormalizable theory by explicit calculations~\cite{NON}.
Including any matter fields into the Lagrangian also yields a
non-renormalizable theory~\cite{Veltman,matter}.
However, including the full set of gravitational field invariants into
the Lagrangian, that is, by treating general relativity as an effective
theory, where the Einstein term provides the lowest energy classical
limit, the action is trivially renormalizable~\cite{Weinberg}.

The effective theory of general relativity provides a mean for
calculating
gravitational quantum corrections, see~\cite{Donoghue:dn,B1,B2}.
The effective field theory approach is therefore at present the
best candidate for a quantum field theory of gravitational interactions in
four dimensions.
The effective field theory will only work for energy scales below the
Planck scale $\sim 10^{19}$ GeV, but this is still incredible good compared
to the limiting scale of the Standard Model $\sim 100-1000$ GeV.
Treating general relativity as an effective field theory is hence the
natural framework for the low energy limit of a gravitational quantum
field theory.

For the theory of Yang-Mills non-abelian vector field theory, the pure
Lagrangian for the non-abelian vector field can be written
\begin{equation}
{\cal L} = -\frac1{8g^2}{\rm tr}{F}^2_{\mu\nu}
\end{equation}
where $A_\mu$ is the vector field and $F_{\mu\nu} = \partial_\mu
A_\nu - \partial_\nu A_\mu + g[A_\mu,A_\nu]$.
The trace in the Lagrangian is over the generators of the non-abelian
Lie-algebra, e.g.
${\rm tr}(T^a T^b T^c)F^a_{\mu\lambda}F^b_{\lambda\nu}F^c_{\nu\mu}$,
where the $T^a$'s are the generators of the
algebra~\footnote{\footnotesize{The notation for the Yang-Mills action
i.e. the $-\frac18\ldots$ follows the non-standard conventions
of~\cite{Tsey1}. Please note that in this letter the action is not
multiplied with $(2\pi\alpha')^2$.}}. The convention $g=1$ will also be
employed, as this is convenient when working with the KLT-relations.
The above non-abelian Yang-Mills action is of course renormalizable
at $D=4$, but it still makes
sense to treat it as the minimal Lagrangian in an effective field theory.
In the modern view of field theory, any invariant obeying the
underlying basic field symmetries, such as local gauge invariance, should
in principle always be allowed into the Lagrangian.

Thus any gauge invariant operator in Yang-Mills and in general relativity is
allowed in the effective actions in this approach.
However, only operators that are non-vanishing on-shell,
and which are invariant under a general local field redefinitions
such as e.g.
\begin{equation}
\delta g_{\mu\nu} = a_1R_{\mu\nu} + a_2g_{\mu\nu}R \ldots
\end{equation}
\begin{equation}
\delta A_\mu = \tilde a_1{\cal D}_\nu F_{\nu\mu} + \ldots
\end{equation}
are unambiguous when comparing on-shell S-matrix
elements~\cite{Tsey1,Tsey2}, where ${\cal D}$ is the gauge theory covariant
derivative. The ellipses in the above relations
stand for possible other field redefinitions, such as higher
derivative and other additional contributions.
Excluding all such terms rules out all contributions containing e.g.
the Ricci tensor, and reduce the possible invariants to include
in the generic effective Lagrangians to a minimum.

Following~\cite{Tsey1} the effective non-abelian Yang-Mills action
for a massless vector field takes the generic form
including all independent invariant field operators
\begin{equation}\begin{split}
{\cal L} &= -\frac18{\rm tr}\Big[F_{\mu\nu}^2 + \alpha'\big(a_1F_{\mu\lambda}F_{\lambda\nu}F_{\nu\mu}
+a_2{\cal D}_\lambda F_{\lambda\mu}{\cal D}_\rho F_{\rho\mu}\big)\\&
+(\alpha')^2\big(a_3F_{\mu\lambda}F_{\nu\lambda}F_{\mu\rho}F_{\nu\rho}
+a_4F_{\mu\lambda}F_{\nu\lambda}F_{\nu\rho}F_{\mu\rho}\\
&+a_5F_{\mu\nu}F_{\mu\nu}F_{\lambda\rho}F_{\lambda\rho}
+a_6F_{\mu\nu}F_{\lambda\rho}F_{\mu\nu}F_{\lambda\rho}\\
&+a_7F_{\mu\nu}{\cal D}_\lambda F_{\mu\nu}{\cal D}_\rho
F_{\rho\lambda} + a_8 {\cal D}_\lambda F_{\lambda \mu}{\cal D}_\rho
F_{\rho\nu}F_{\mu\nu} \\
&+a_9{\cal D}_\rho{\cal D}_\lambda F_{\lambda\mu}{\cal D}_\rho {\cal D}_\sigma F_{\sigma\mu}\big)+\ldots \Big]
\end{split}\end{equation}
where $\alpha'$ is the string tension,
$F_{\mu\nu} = \partial_\mu A_\nu - \partial_\nu A_\mu + [A_\mu,A_\nu]$
and ${\cal D}$ is the non-abelian covariant derivative.

Excluding all but the on-shell non-vanishing and local field transformation
invariant terms, one arrives at
\begin{equation}\begin{split}
{\cal L} &= -\frac18{\rm tr}\Big[F_{\mu\nu}^2 + \alpha'\big(a_1 F_{\mu\lambda}F_{\lambda\nu}F_{\nu\mu}\big)\\&
+(\alpha')^2\big(a_3F_{\mu\lambda}F_{\nu\lambda}F_{\mu\rho}F_{\nu\rho}
+a_4F_{\mu\lambda}F_{\nu\lambda}F_{\nu\rho}F_{\mu\rho}\\
&+a_5F_{\mu\nu}F_{\mu\nu}F_{\lambda\rho}F_{\lambda\rho}
+a_6F_{\mu\nu}F_{\lambda\rho}F_{\mu\nu}F_{\lambda\rho}\big)+\ldots
\Big]\end{split}\end{equation}

The general Lagrangian will be written separately for left and right
non-interacting fields,
with generically different left-right coefficients.
So for, e.g., the left Lagrangian one has
\begin{equation}\begin{split}\label{eq7}
{\cal L}^L_{YM} &= -\frac18{\rm tr}\Big[F^L_{\mu\nu}F^L_{\mu\nu}   +
\alpha'\big(a^L_1 F^L_{\mu\lambda}F^L_{\lambda\nu}F^L_{\nu\mu}\big)\\&
+(\alpha')^2\big(a^L_3F^L_{\mu\lambda}F^L_{\nu\lambda}F^L_{\mu\rho}F^L_{\nu\rho}+a^L_4F^L_{\mu\lambda}F^L_{\nu\lambda}F^L_{\nu\rho}F^L_{\mu\rho}\\
&+a^L_5F^L_{\mu\nu}F^L_{\mu\nu}F^L_{\lambda\rho}F^L_{\lambda\rho}
+a^L_6F^L_{\mu\nu}F^L_{\lambda\rho}F^L_{\mu\nu}F^L_{\lambda\rho}\big)
+\ldots \Big]\end{split}\end{equation}
The expression for the right Lagrangian is identical, but with everywhere $F^L_{\mu\nu}\rightarrow F^R_{\mu\nu}$ and
e.g. $a_1^L\rightarrow a_1^R$ etc.

The inclusion of independent left and right field invariants is important
because the KLT-relations allow for different left and right scattering
amplitudes. In order to include string solutions such as heterotic
strings, the possibility for a generally different left and right
scattering amplitude has to be included into the formalism. Heterotic
string solutions will however not be considered explicitly here.

In general relativity a similar situation occurs for the effective
action, following~\cite{Tsey2}
the most general Lagrangian one can consider is
\begin{equation}\begin{split}
{\cal L} &= \frac{2\sqrt{-g}}{\kappa^2}\Big[R + \alpha'\big (a_1
R_{\lambda\mu\nu\rho}^2 + a_2R_{\mu\nu}^2 +a_3R^2\big)\\&
+(\alpha')^2\big(b_1 R^{\mu\nu}_{\ \  \alpha\beta}
R^{\alpha\beta}_{\ \  \lambda\rho} R^{\lambda\rho}_{\ \
\mu\nu}+b_2(R^{\mu\nu}_{\ \  \alpha\beta} R^{\alpha\beta}_{\ \
\lambda\rho} R^{\lambda\rho}_{\ \  \mu\nu}\\&-2R^{\mu\nu\alpha}_{\
\ \ \beta} R_{\ \ \nu\lambda}^{\beta\gamma} R^{\lambda}_{\ \mu
\gamma\alpha}) + b_3 R_{\mu\alpha\beta\gamma}
R^{\alpha\beta\gamma}_{\ \ \ \ \nu}R^{\mu\nu} + b_4
R_{\mu\nu\rho\lambda}R^{\nu\lambda}R^{\mu\rho}\\&+b_5R_{\mu\nu}R^{\nu\lambda}R^\mu_{\
\lambda} +b_6R_{\mu\nu}{\cal
D}^2R^{\mu\nu}+b_7R^2_{\lambda\mu\nu\rho}R
\\&+b_8R_{\mu\nu}^2R+b_9R^3+b_{10}R{\cal D}^2R\big)+\ldots\Big]
\end{split}\end{equation}
again most terms can be removed be a field redefinition. Here
$\cal D$ denotes the gravitational covariant derivative.
This leaves only the terms
\begin{equation}\begin{split}\label{eq9}
{\cal L} &= \frac{2\sqrt{-g}}{\kappa^2}\Big[R + \alpha'\big(a_1 R_{\lambda\mu\nu\rho}^2\big)
+(\alpha')^2\big(b_1 R^{\mu\nu}_{\ \  \alpha\beta}
R^{\alpha\beta}_{\ \  \lambda\rho}
R^{\lambda\rho}_{\ \  \mu\nu}\\& +b_2 (R^{\mu\nu}_{\ \  \alpha\beta}
R^{\alpha\beta}_{\ \  \lambda\rho}
R^{\lambda\rho}_{\ \  \mu\nu}-2R^{\mu\nu\alpha}_{\ \ \ \beta}
R_{\ \ \nu\lambda}^{\beta\gamma}
R^{\lambda}_{\ \mu\gamma\alpha})\big)+\ldots\Big]
\end{split}\end{equation}
to be considered when comparing S-matrix elements.
In the above form of the field redefinition
invariant effective action the conventions of~\cite{Tsey2} is followed.

\section{Scattering amplitudes in Yang-Mills theory and gravity}
In this section we check and review results contained already in
refs.~\cite{Tsey1,Tsey2}, for use in later sections.
In order to calculate the scattering amplitudes from the effective
actions, one must expand the gravity Lagrangian.
The 3-point amplitudes in gravity is rather uncomplicated, as one can exclude
all terms that vanish on-shell. The on-shell non-vanishing
contributions in the effective gravitational action, can then be found as
\begin{equation}
R=-\frac{\kappa^3}4h_{\alpha\beta}(h_{\lambda\rho}\partial_\alpha\partial_\beta h_{\lambda\rho}+
2\partial_\alpha h_{\lambda\rho}\partial_\rho h_{\beta\lambda})+\ldots,
\end{equation}
\begin{equation}
R^2_{\lambda\mu\nu\rho}-4R^2_{\mu\nu}+R^2 = -\kappa^3h_{\mu\nu}\partial_\mu\partial_\rho h_{\alpha\beta}\partial_\alpha\partial_\beta h_{\nu\rho}+\ldots,
\end{equation}
\begin{equation}
R^{\mu\nu}_{\ \ \alpha\beta}R^{\alpha\beta}_{\ \ \lambda\rho}R^{\lambda\rho}_{\ \ \mu\nu} =
\kappa^3\partial_\rho \partial_\alpha h_{\beta\nu}\partial_\beta\partial_\lambda h_{\mu\alpha}\partial_\mu\partial_\nu h_{\lambda\rho}+\ldots
\end{equation}
Here $g_{\mu\nu}\equiv\eta_{\mu\nu}+\kappa h_{\mu\nu}$,
where $\eta_{\mu\nu}$ is the flat metric. Because
of the flat background metric
no distinctions are made of up and down indices on the right hand
side of the equations, so indices can be put where convenient.
The results are identical to these of ref.~\cite{Tsey2}, and gives the
following generic 3-point scattering amplitude for the effective
action of general relativity
\begin{equation}\begin{split}
{M_3}&= \kappa \Big[
\zeta_2^{\mu\sigma}\zeta_3^{\mu\rho}
\big(\zeta_1^{\alpha\beta}k_2^{\alpha}k_2^{\beta}\delta^{\sigma\rho}
+\zeta_1^{\sigma\alpha}k_2^{\alpha}k_1^{\rho}
+\zeta_1^{\sigma\alpha}k_3^{\alpha}k_1^{\rho}\big)\\&
+\zeta_1^{\mu\sigma}\zeta_3^{\mu\rho}
\big(\zeta_2^{\alpha\beta}k_3^{\alpha}k_3^{\beta}\delta^{\sigma\rho}
+\zeta_2^{\sigma\alpha}k_1^{\alpha}k_2^{\rho}
+\zeta_2^{\sigma\alpha}k_3^{\alpha}k_2^{\rho}\big)\\&
+\zeta_1^{\mu\sigma}\zeta_2^{\mu\rho}
\big(\zeta_3^{\alpha\beta}k_1^{\alpha}k_1^{\beta}\delta^{\sigma\rho}
+\zeta_3^{\sigma\alpha}k_1^{\alpha}k_3^{\rho}
+\zeta_3^{\sigma\alpha}k_2^{\alpha}k_3^{\rho}\big)\\&
+\alpha'[4a_1\zeta_2^{\mu\sigma}\zeta_3^{\mu\rho}\zeta_1^{\alpha\beta}k_2^\alpha k_2^\beta k_3^\sigma k_2^\rho\\&
+4a_1\zeta_1^{\mu\sigma}\zeta_3^{\mu\rho}\zeta_2^{\alpha\beta}k_3^\alpha k_3^\beta k_2^\sigma k_1^\rho
+4a_1\zeta_1^{\mu\sigma}\zeta_2^{\mu\rho}\zeta_3^{\alpha\beta}k_1^\alpha k_1^\beta k_2^\sigma k_3^\rho]
\\&+(\alpha')^2[12b_1\zeta_1^{\alpha\beta}\zeta_2^{\gamma\delta}\zeta_3^{\tau\rho}k_1^\tau k_1^\rho
k_2^\alpha k_2^\beta k_3^\gamma k_3^\delta]
\Big]
\end{split}\end{equation}
where $\zeta_i^{\mu\nu}$ and $k_i$, $i=1,..,3$ denote the polarization
tensors and momenta for the external graviton legs.

In the Yang-Mills case the following result for the generic 3-point
amplitude can be worked out. Here it is written for e.g. the left vector
fields
\begin{equation}\begin{split}
A_{3L} &= -\Big[(\zeta_3\cdot k_1 \zeta_1 \cdot \zeta_2 + \zeta_2\cdot k_3 \zeta_3 \cdot \zeta_1+\zeta_1\cdot k_2 \zeta_2 \cdot \zeta_3)
\\&+ \frac34\alpha'a_1^L\zeta_1\cdot k_2 \zeta_2\cdot k_3 \zeta_3 \cdot k_1\Big]
\end{split}\end{equation}
This result is equivalent to that of ref.~\cite{Tsey1}. With a similar
notation as used above in the gravity case, $\zeta_i^\mu$, $k_i$,
$i=1,..,3$ denotes the polarizations and momenta for the external
vector lines.
The on-shell 4-point amplitude is generically a rather complicated
expression, as many terms are non-vanishing on-shell.
But in order to match the amplitudes of the gauge theory with
gravity, only certain specific parts of the
amplitude need to be compared~\cite{Tsey1,Tsey2,Deser,shimada}.
Gauge invariance will dictate the remaining parts of the amplitude
to go along.
Using the part of the scattering amplitude where no momenta is
contracted with an external polarization index, simplifies the
calculations. Besides the 4-point contact vertices there are three
different types of graviton exchanges to order $\alpha'^2$, these
contributions have
to be included too in the 4-point amplitude. The 3-point parts giving
these contributions i.e.  with one leg off-shell and the two other
satisfying the on-shell external polarization index constraint can
be found below.
Expanding the Lagrangian, one finds the following contributions
\begin{equation}
R=-\frac{\kappa^3}4h_{\mu\nu}h_{\nu\rho}\partial^2h_{\rho\mu}+\ldots,
\end{equation}
\begin{equation}
R^2_{\lambda\mu\nu\rho}-4R^2_{\mu\nu}+R^2 = -\frac{\kappa^3}2\partial_\alpha h_{\mu\nu}\partial_\alpha h_{\nu\rho}\partial^2 h_{\rho\mu}+\ldots,
\end{equation}
\begin{equation}\begin{split}
R^{\mu\nu}_{\ \ \alpha\beta}R^{\alpha\beta}_{\ \ \lambda\rho}R^{\lambda\rho}_{\ \ \mu\nu} &=
-\frac{3\kappa^3}2\partial_\alpha\partial_\beta h_{\mu\nu} \partial_\alpha\partial_\beta h_{\nu\rho}\partial^2 h_{\rho\mu}
\\&+3\kappa^4(h_{\mu\nu}\partial_\alpha\partial_\beta h_{\nu\rho} \partial_\alpha\partial_\gamma h_{\rho\sigma} \partial_\beta\partial_\gamma h_{\sigma\mu}
\\&+\frac{1}2\partial_\alpha h_{\mu\nu}\partial_\beta h_{\nu\rho} \partial_\alpha\partial_\gamma h_{\rho\sigma}
\partial_\beta\partial_\gamma h_{\sigma\mu})+\ldots,
\end{split}\end{equation}
\begin{equation}\begin{split}
R^{\mu\nu\alpha}_{\ \ \ \beta}R_{\ \ \nu\lambda}^{\beta\gamma}R^{\lambda}_{\ \mu\gamma\alpha}&=
-\frac{3\kappa^4}8\partial_\alpha h_{\mu\nu}\partial_\beta h_{\nu\rho} \partial_\beta \partial_\gamma h_{\rho\sigma} \partial_\alpha\partial_\gamma h_{\sigma\mu}+\ldots,
\end{split}\end{equation}
The same conventions as for 3-point terms are used in these equations.
The above results simply verify ref.~\cite{Tsey2}.

The 4-point gravity scattering amplitude can then be found to be~\cite{Tsey2}
\begin{equation}\begin{split}
{M_4} &= \frac12
\frac{\kappa^2}{\alpha'}\Big[\zeta_1\zeta_2\zeta_3\zeta_4(z + a_1
z^2 -3b_1(z^3-4xyz) \\& -3b_2(z^3-\frac72xyz) +(a_1^2+3b_1 +
3b_2)(z^3-3xyz))\\& + \zeta_1\zeta_2\zeta_4\zeta_3(y + a_1 y^2
-3b_1(y^3-4xyz)\\&-3b_2(y^3-\frac72xyz) +(a_1^2+3b_1 +
3b_2)(y^3-3xyz))\\& + \zeta_1\zeta_3\zeta_2\zeta_4(x + a_1 x^2
-3b_1(x^3-4xyz)\\&-3b_2(x^3-\frac72xyz) +(a_1^2+3b_1 +
3b_2)(x^3-3xyz))\Big]
\end{split}\end{equation}
where $\zeta_1\zeta_2\zeta_3\zeta_4 = \zeta_1^{\alpha\beta}\zeta_2^{\beta\gamma}\zeta_3^{\gamma\delta}\zeta_4^{\delta\alpha}$ and $x =
-2\alpha'(k_1\cdot k_2)$,
$y = -2\alpha'(k_1\cdot k_4)$ and $z=-2\alpha'(k_1\cdot k_3)$.

In the Yang-Mills case the corresponding left scattering amplitude reads
\begin{equation}\begin{split}
A_{4L} &= \bigg[
\Big[\zeta_{1324}+\frac{z}{x}\zeta_{1234}+\frac{z}{y}\zeta_{1423}\Big]
\\&+\Big[-\frac{3a^L_1}{8}z(\zeta_{1324}+\zeta_{1234}+\zeta_{1423})\Big]
\\&+\Big[\frac{9(a^L_1)^2}{128}(x(z-y)\zeta_{1234}+y(z-x)\zeta_{1423})\Big]\\&
-\frac14\Big[(\frac12a^L_3)xy\zeta_{1324}-(\frac14a_3^L+2a_6^L)z^2\zeta_{1324}
\\&+(\frac14a^L_3 +\frac12a^L_4)yz\zeta_{1234} \\& + (\frac14a^L_3 + a^L_5)zx\zeta_{1234}+(\frac12a_4^L+a_5^L)yx\zeta_{1234}
\\&+(\frac14a^L_3 +a_5^L)yz\zeta_{1423}\\&+(\frac14a_3^L+\frac12a_4^L)xz\zeta_{1423}+(\frac12a_4^L+a_5^L)xy\zeta_{1423}\Big]\bigg]
\end{split}\end{equation}
where $x$, $y$ and $z$ is defined as above, and e.g.
$\zeta_{1234} = (\zeta_1 \cdot \zeta_2)(\zeta_3 \cdot \zeta_4)$
and etc. Again we look only at the part
where no polarization index is contracted with a momentum index.
The contact terms in the above expression were calculated explicitly
and the results agree with those of ref.~\cite{Tsey2}.
The non-contact terms were adapted from~\cite{Tsey2}.

\section{The open-closed string relations}
The relations between open and closed string dictates that the
general $M$-particle function of the closed string is related to
a product of open strings in the following manner.
The conventions of~\cite{GSW} are here followed and the relation is written as
\begin{equation}\begin{split}
{A}^M_{\rm closed} \sim \sum_{\Pi,\tilde \Pi}
e^{i\pi\Phi(\Pi,\tilde\Pi)}{A}_M^\text{left open}(\Pi)
{A}_M^\text{right open}(\tilde \Pi)
\end{split}\end{equation}
where $\Pi$ and $\tilde\Pi$ are particular cyclic orderings of the
open-string external lines associated with the right and left moving modes.
The function $\Phi(\Pi,\tilde \Pi)$ is the appropriate phase factor of the
exponential associated with the explicit cyclic permutations. This
open-closed string relation has been derived from string theory, but it
holds as well in field theory.

For the 3- and 4-point amplitudes, the following specific KLT-relations
are adapted~\footnote{\footnotesize{The specific
forms of the KLT-relations do not follow any specific convention e.g.~\cite{Bern4}.
In order the keep the conventions of~\cite{Tsey1,Tsey2}, the employed
relations are normalized
to be consistent for the O(1) terms of the 3- and 4-point scattering
amplitudes.}}
\begin{equation}\begin{split}\label{eq22}
{M}_{\rm 3 \ gravity}^{\mu\tilde\mu\nu\tilde\nu\rho\tilde\rho}&(1,2,3) = \kappa{A}^{\mu\nu\rho}_\text{3 L-gauge}(1,2,3)\times {A}^{\tilde\mu\tilde\nu\tilde\rho}_\text{3 R-gauge}(1,2,3)
\end{split}\end{equation}
and
\begin{equation}\begin{split}\label{eq23}
{M}_{\rm 4 \
gravity}^{\mu\tilde\mu\nu\tilde\nu\rho\tilde\rho\sigma\tilde\sigma}&(1,2,3,4)
= \frac{\kappa^2}{4\pi\alpha'}\sin(\pi x)\\&\times
{A}^{\mu\nu\rho\sigma}_\text{4
L-gauge}(1,2,3,4)\times{A}^{\tilde\mu\tilde\nu\tilde\rho\tilde\sigma}_\text{4
R-gauge}(1,2,4,3)
\end{split}\end{equation}
where $M$ is a gravity tree amplitude, and $A$ is the color ordered
amplitude for the gauge theory with coupling constant $g=1$.
As previously the definition $x = -2\alpha'(k_1\cdot k_2), \ldots$
is employed,
where $k_1$ and $k_2$ are particular momenta of the external lines.

\section{Mapping operators via the KLT-relations}
We now present the main results of this paper linking the
general relativity operators to the Yang-Mills ones via the
KLT-relations. This is done by demanding that the scattering amplitudes
generated by the Yang-Mills effective Lagrangian in eq. (\ref{eq7}) (and the
corresponding 'R' one) and the general relativity effective Lagrangian
in eq. (\ref{eq9}) satisfy the KLT-relations eq. (\ref{eq22}) and eq. (\ref{eq23}). In this
way, the following equations are found to hold for the generic coefficients in the effective
actions:
\begin{equation}\begin{split}
\frac{3a_1^L}{16}+\frac{3a_1^R}{16}&=a_1,\\
\frac{3a_1^L\, a_1^R}{64} &= b_1,\\
6a_5^L+3a_4^L+\frac{27(a_1^L)^2}{16} &= 0,\\
6a_5^R+3a_4^R+\frac{27(a_1^R)^2}{16} &= 0,
\end{split}\end{equation}
together with
\begin{equation}\begin{split}
96a_1^2&= 6a_3^L+3a_3^R+18a_4^L+12a_5^L+24a_6^R\\&+\frac{81(a_1^L)^2}{8}-16\pi^2,\\
96a_1^2&= 3a_3^L+6a_3^R+18a_4^R+12a_5^R+24a_6^L\\&+\frac{81(a_1^R)^2}{8}-16\pi^2,\\
96a_1^2&= 6a_3^L-3a_3^R-6a_4^L-36a_5^L-12a_5^R-16\pi^2\\&-\frac{27(a_1^L)^2}{8}+\frac{27(a_1^R)^2}{8},\\
96a_1^2&=-3a_3^L+6a_3^R-6a_4^R-12a_5^L-36a_5^R-16\pi^2\\&+\frac{27(a_1^L)^2}{8}-\frac{27(a_1^R)^2}{8},
\end{split}\end{equation}
\begin{equation}\begin{split}
96b_1+48b_2&=4a_3^L+5a_3^R+2a_4^L+6a_4^R+8a_6^L\\&
+\frac{9(a_1^R)^2}{4}+\frac{9a_1^L\, a_1^R}{2}-16\pi^2,\\
96b_1+48b_2&=4a_3^R+5a_3^L+2a_4^R+6a_4^L+8a_6^R\\&
+\frac{9(a_1^L)^2}{4}+\frac{9a_1^L\, a_1^R}{2}-16\pi^2,
\end{split}\end{equation}
\begin{equation}\begin{split}
96a_1^2-96b_1-48b_2& = a_3^L+2a_4^R-12a_5^L-12a_5^R\\&-\frac{32\pi^2}3-\frac{9a_1^L\, a_1^R}{2}+\frac{9(a_1^L)^2}{8}+\frac{9(a_1^R)^2}{8},\\
96a_1^2-96b_1-48b_2& = 2a_4^R-4a_5^L-12a_5^R+8a_6^L\\&-\frac{16\pi^2}3-\frac{9a_1^L\, a_1^R}{2}+\frac{9(a_1^L)^2}{8}+\frac{9(a_1^R)^2}{8},\\
96a_1^2-96b_1-48b_2& = 2a_4^L-12a_5^L-4a_5^R+8a_6^R\\&-\frac{16\pi^2}3-\frac{9a_1^L\, a_1^R}{2}+\frac{9(a_1^L)^2}{8}+\frac{9(a_1^R)^2}{8},\\
96a_1^2-96b_1-48b_2& =
a_3^R+2a_4^L-12a_5^L-12a_5^R\\&-\frac{32\pi^2}3-\frac{9a_1^L\,
a_1^R}{2}+\frac{9(a_1^L)^2}{8}+\frac{9(a_1^R)^2}{8}.
\end{split}\end{equation}

Solving these equations, one ends up with the following set of relations between the coefficients of the effective gravitational Lagrangian
and the Yang-Mills vector Lagrangian:
\begin{equation}\begin{split}
a_1^L&=a_1^R=\frac8{3}a_1,\\
a_3^L&=a_3^R=\frac{4\pi^2}3,\\
a_4^L&=a_4^R=\frac{2\pi^2}3,\\
a_5^L&=a_5^R=-\frac{\pi^2}{3}-2a_1^2,\\
a_6^L&=a_6^R=-\frac{\pi^2}{6}+2a_1^2,\\
b_1&=\frac{1}{3}a_1^2,\\
b_2&=\frac{2}{3}a_1^2
\end{split}\end{equation}
Thus we see that in order for there to be a mapping at the order $\alpha'^2$,
eq. (\ref{eq9}) must take on particular values.
As seen from the above solution the coefficients $a^L_3$, $a^R_3$ and $a^L_4$,
$a^R_4$, are completely fixed by the KLT-relations for a given effective
gravity Lagrangian satisfying the constraints on $b_1$ and $b_2$.
Thus the KLT-relations are rather constraining on the form of
the general relativity Lagrangians that can be mapped into gauge theory.

This letter considers terms to order $O(\alpha'^3)$ in the
effective Lagrangians
and have dealt with 3- and 4-point amplitudes.
Investigation of the mapping of operators between gravity
and Yang-Mills theory can be carried out at higher orders of
$\alpha'$ and will possibly lead to further insight of the
mapping process.

\section{Discussion}
The solution space of the above equations suggest the following
interpretation.
For a given coefficient $a_1$ in the gravitational effective field
theory action through order $\alpha'^2$, there is a set of unique
(left-right) effective Yang-Mills Lagrangians,
up to a field redefinition, that satisfies
the constraint of the KLT-relations. The set of solutions
span out a space of operator solutions, where the different
types of strings correspond to certain points along the path.
For string theories with similar left and right Lagrangians,
one can find a bosonic string corresponding to the
solution: $a_1^L = a_1^R = \frac83$,
$a_3^L=a_3^R=\frac{4\pi^2}3$, $a_4^L=a_4^R=\frac{3\pi^2}2$,
$a_5^L=a_5^R=-\frac{\pi^2}{3}-2$,
and $a_6^L=a_6^R=-\frac{\pi^2}{6}+2$, on the open string
side, and $a_1=1$, $b_1=\frac13$ and $b_2 = \frac23$, on the closed
string side, while the superstring correspond to:
$a_1^L = a_1^R =0$, $a_3^L=a_3^R=\frac{4\pi^2}3$,
$a_4^L=a_4^R=\frac{2\pi^2}3$, $a_5^L=a_5^R=-\frac{\pi^2}{3}$
and $a_6^L=a_6^R=-\frac{\pi^2}{6}$, for the open string,
and $a_1=b_1=b_2=0$ for the closed string.
Also the heterotic string can be fitted to the above solution space
but here a Chern-Simons term is needed in the effective action, and
the left and right coefficients in the action will be dissimilar
and correspond respectively to a bosonic and a superstring solution.

Since $a_1$ are arbitrary in the above equations, the solution space
of the amplitude matching is much larger than the traditional set
of string solutions. That is, non-traditional string solutions are
allowed by the KLT-relations.

The assumption of the KLT-relations provide us with a very
detailed and constraining solution mapping operators of general relativity
into corresponding
non-abelian Yang-Mills operators. The $R_{\mu\nu\alpha\beta}^2$ term of
the effective action of general relativity is directly seen to correspond
to vertex products of a ${{\rm tr} F_{\mu\nu}^2}$ term and a
${{\rm tr} F_{\mu\nu}^3}$ term
of the non-abelian Yang-Mills effective action. The mapping of the
higher order terms is more complicated, but still it is no problem
to find the solution.
It appears {\it a priori} rather surprising that an explicit mapping process
is possible for such sets of dissimilar operators.

The results for the string solutions are identical to the results
of~\cite{Tsey1,Tsey2,Tsey3}, but the viewpoint here is very different.
When one matches the coefficients of the effective Lagrangian to the
traditional string solutions, only such solutions appear. Matching
the effective
Lagrangians through the KLT-relations leaves a broader solution
space, where non-traditional string solutions are allowed too.
The effective actions provide full and very general theories
both for Yang-Mills and for gravity, at energy scales up to the Planck scale.
Furthermore the link from the effective action to
the low-energy Lagrangian is obvious. This is not the case in string
theory where a Lagrangian formalism is not natural.
String theory in itself is found not to be required for the mapping, only the
KLT-relations which hold more generally.
Still it is hard to see how the KLT-relationship would have been
found without string theory, and furthermore it is very useful
to know the string solutions, as a guide in the mapping process.

Only the pure gravity and Yang-Mills effective actions have been discussed in this letter,
but clearly matter sources could also be included. This would introduce
new aspects in the mapping process, and additional knowledge of the mapping of
operators could possibly be extracted. Another starting point for
further investigations is
perhaps to look upon relaxed forms of the KLT-relations.
That is: allowing for
more general mappings of the operators, e.g. replacing
the $\sin(\pi x)$ with a general series in $\pi x$. The possibility
of a Chern-Simons term
in the effective gravity action and the heterotic string solution
is yet another issue.

At the present stage the KLT-relations point out some clues of non-trivial
links between gauge theory and gravity, but still no full understanding of
this has been extracted. Clearly future investigations should focus on
this problem.

\begin{acknowledgments}
I would like to thank Department of physics, UCLA, for its very
kind hospitality and to express my gratitude towards Zvi Bern for
his invaluable help during my stay at UCLA.
I would like to thank him for suggesting to use the
KLT-relations and the links between gauge theory and gravity and for
his reading of the manuscript. I would also like to
thank P. H. Damgaard for discussions.
\end{acknowledgments}

\end{document}